\documentclass[12pt]{article}

\usepackage[T1]{fontenc}
\usepackage[utf8]{inputenc}
\usepackage[italian, english]{babel}
\usepackage[top=1in, bottom=1in, left=1.25in, right=1.25in]{geometry}
\usepackage{amsmath, amssymb, amsthm, amscd, amstext, amsxtra}
\usepackage{graphicx, hyperref, lineno, natbib}
\usepackage{bibentry}
\usepackage[dvipsnames, svgnames, x11names]{xcolor}
\usepackage{multirow, arydshln}
\usepackage[plain,noend]{algorithm2e}

\graphicspath{{./figures/}}

\newcommand{\Exp}{\mathrm{Exp}}
\newcommand{\Dir}{\mathrm{Dir}}
\newcommand{\Ga}{\mathrm{Ga}}

\newcommand{\iG}{\mathrm{iG}}
\newcommand{\gig}{\mathrm{GiG}}
\newcommand{\N}{\mathcal{N}}

\newcommand{\thetac}{\theta_0}
\newcommand{\psic}{\psi_0}
\newcommand{\tauc}{\tau_0}
\newcommand{\phic}{\phi_0}
\newcommand{\thetan}{\theta^\prime}
\newcommand{\psin}{\psi^\prime}
\newcommand{\taun}{\tau^\prime}
\newcommand{\phin}{\phi^\prime}
\newcommand{\rpkg}[2]{\textit{#1}\nocite{#2}}

\newtheorem{proposition}{Proposition}

\title{On the Posterior Computation Under the Dirichlet-Laplace Prior}
\author{ P. Onorati\textsuperscript{1}, D.B. Dunson \textsuperscript{2}, and A. Canale\textsuperscript{1}
	\\
	\small{\textsuperscript{1}Department of Statistical Sciences, University of
Padova, Italy} 
  \\
  \small{\textsuperscript{2}Department of Statistical Science, Duke University
  } }
\date{}

\begin{document}

\maketitle

\begin{abstract}
  Modern applications routinely collect high-dimensional data, leading to
  statistical models having more parameters than there are samples available. A
  common solution is to impose sparsity in parameter estimation, often using
  penalized optimization methods. Bayesian approaches provide a probabilistic
  framework to formally quantify uncertainty through shrinkage priors. Among
  these, the Dirichlet-Laplace prior has attained recognition for its
  theoretical guarantees and wide applicability. This article identifies a
  critical yet overlooked issue in the implementation of Gibbs sampling
  algorithms for such priors. We demonstrate that ambiguities in the
  presentation of key algorithmic steps, while mathematically coherent, have led
  to widespread implementation inaccuracies that fail to target the intended
  posterior distribution---a target endowed with rigorous asymptotic guarantees.
  Using the normal-means problem and high-dimensional linear regressions as
  canonical examples, we clarify these implementation pitfalls and their
  practical consequences and propose corrected and more efficient sampling
  procedures.
\end{abstract}

\begin{center}
\end{center}

\section{Introduction} \label{sec:intro} 

High-dimensional parameter spaces have become commonplace in modern statistics
and machine learning. Often, the dimensionality of the parameter space far
exceeds the available information encapsulated within the observations. In these
settings, it is typical to assume that  the underlying signal resides within a
sparse subset of the parameter space, with many parameters vanishing to zero.
This parsimonious perspective spread in the early 90's boosted by the successful
$L_1$/LASSO penalization \citep{tibshirani1996, efron2004} and its many
generalizations \citep{enet,glasso,adalasso}. 

Notably, there exists an equivalence between penalized optimization and Bayesian
maximum a posteriori estimation. For example, the LASSO implicitly assumes
double exponential (Laplace) priors. Yet, the Bayesian framework extends far
beyond maximum a posteriori estimation, offering a probabilistic mechanism to
fully quantify uncertainty in an elegant way. Bayesian shrinkage priors can be
broadly divided into spike-and-slab formulations
\citep{mitchell1988,george1993,ishwaran2005}, which explicitly mix a point mass
at zero with a diffuse component, continuous-shrinkage priors
\citep{carvalho2010,armagan2013}, which avoid Dirac spikes by concentrating
density near zero while preserving heavy tails, and hybrid solutions
\citep{rockova2018,rockova2018a}. 

Within the class of continuous shrinkage methods, the Dirichlet-Laplace prior,
proposed by \citet{bhattacharya2015}, has emerged as a prominent framework in
the last ten years. This approach has garnered considerable attention for its
strong theoretical guarantees, including optimal posterior concentration rates
under specific regimes, along with broad practical utility. The
Dirichlet-Laplace prior is distinguished by its adaptability, serving as a
default choice in diverse scenarios. \citet{wei2020} established optimal
concentration properties for shrinkage priors, including the Dirichlet-Laplace
prior, in logistic regression settings. Further extensions of the framework have
been developed for linear regression \citep{zhang2018} and generalized linear
models \citep{bhattacharyya2022,bhattacharyya2022a}.

This article focuses on the posterior computation under the Dirichlet-Laplace
prior. We highlight a subtle, yet critical aspect of the Gibbs sampling
algorithm as presented by \citet{bhattacharya2015}. Specifically,  certain
procedural details, while mathematically coherent, may inadvertently lead to
wrong implementation of the code, thus producing algorithms not targeting the
true posterior distribution. This ambiguity propagated to generalizations and,
for example, is also present in \citet{zhang2018} for the case of the linear
regression model. We posit that this oversight has led to widespread
implementation inaccuracies in applied work, including popular computational
tools such as the dlbayes R package \citep{dlbayes}.

Before carefully discussing the issue, we introduce the framework and the
notation. Let $\N(\mu, \sigma^2)$ denote the Gaussian distribution with mean
$\mu$ and variance $\sigma^2$, $\Exp(\lambda)$ denote an exponential
distribution with rate parameter $\lambda$, $\Ga(\nu, \lambda)$ denote a gamma
distribution with shape $\nu$ and rate $\lambda$, and  $\Dir_d(\alpha_1, \dots,
\alpha_d)$ denote a $d$-dimensional Dirichlet distribution with concentration
parameter vector $(\alpha_1, \dots, \alpha_d)^\top$. Then, the Dirichlet-Laplace
prior for a $p$-dimensional problem  is defined as
\begin{align}
  \theta_j \mid \psi_j, \tau, \phi_j &\sim \N(0, \psi_j \tau^2 \phi^2_j) \, , \,
  \psi_j \sim \Exp(1 / 2) \, , \, j = 1, \dots, p \, , \nonumber
  \\
  \tau &\sim \Ga(p a, 1 / 2)\, , \, \phi_1, \dots, \phi_p \sim \Dir_p(a, \dots,
  a) \, . \label{eq:DirLapl}
\end{align}
In the Dirichlet-Laplace prior, the components of the concentration vector must
be equal. The common value, denoted with $a$, is the only free hyper-parameter
of the Dirichlet-Laplace prior and controls the amount of the shrinkage. In the
following, let $\iG(\mu, \lambda)$ denote the inverse Gaussian distribution with
mean $\mu$ and shape $\lambda$ and  $\gig(\nu, \gamma, \xi)$ denote a
generalized inverse Gaussian distribution whose density, at $x$, is proportional
to $x^{\nu - 1} \exp \{-(\gamma x + \xi / x) / 2\}$.

\section{Posterior Computation}
\label{sec:dirlapl}

\subsection{Normal Means Problem}
\label{subsec::nmp}

We first   focus on the estimation of  an $n$-dimensional mean based on a single
observation corrupted with independent standard Gaussian noise
\citep{donoho1992maximum,johnstone2004needles,castillo2012needles}, i.e. 
\begin{equation}
  \label{eq:nmp}
  y_i = \theta_i + \varepsilon_i , \quad \varepsilon_i \overset{i.i.d.}{\sim}
  \N(0, 1), \quad  i = 1, \dots, n \, .
\end{equation}

Let $p(\theta, \psi, \tau, \phi \mid y)$ be the posterior distribution of the
parameters $\theta, \psi, \tau$, and  $\phi$ given the sample $y = (y_1, \dots,
y_n)^\top$ and the prior reported in \eqref{eq:DirLapl} where the parameter
dimension $p$ is here equal to the number of observations $n$ and the index $i$
has been used in placer of $j$ as commonly done in this context.  
The algorithm presented in Section 2.4 of \citet{bhattacharya2015} samples
iteratively from $p(\theta \mid \psi, \tau, \phi, y)$ and $p(\psi, \tau, \phi
\mid \theta)$. The latter full conditional distribution exploits the
independence of $\psi, \tau, \phi$ from $y$ when they are conditioned on
$\theta$.  The following Algorithm \ref{alg:original} shows how the steps were
reported in \citet{bhattacharya2015}. 
\begin{algorithm}
  \label{alg:original}
  \citet{bhattacharya2015} Gibbs sampler
  \begin{tabbing}
    \qquad \enspace \textit{Step} 1. Sample $\theta \mid \psi, \tau, \phi, y$
    \\
    \qquad \enspace\qquad\qquad For $i=1$ to $i=n$
    \\
    \qquad \qquad\qquad\qquad Set $\zeta^2_i = \{1 + 1 / (\psi_i \tau^2 \phi^2_i
    )\}^{-1}$
    \\
    \qquad \qquad\qquad\qquad Sample $\theta_i \mid \psi_i, \tau, \phi_i, y_i
    \sim \N(\zeta^2_i y_i, \zeta^2_i)$
    \\
    \qquad \enspace \textit{Step} 2. Sample $\psi \mid \theta, \tau, \phi$
    \\
    \qquad \enspace\qquad\qquad For $i=1$ to $i=n$
    \\
    \qquad \qquad\qquad\qquad Sample $\widetilde{\psi}_i \mid \theta_i, \tau, 
    \phi_i, \sim \iG(\tau \phi_i / |\theta_i|, 1)$ and set $\psi_i = 1 / \widetilde{\psi}_i$ 
    \\
    \qquad \enspace \textit{Step} 3. Sample $\tau \mid \theta, \phi, \sim
    \gig(n(a-1), 1, 2 \sum_{i=1}^n |\theta_i| / \phi_i)$
    \\
    \qquad \enspace \textit{Step} 4. Sample $\phi \mid \theta$
    \\
    \qquad \enspace\qquad\qquad For $i=1$ to $i=n$
    \\
    \qquad \qquad\qquad\qquad Sample $T_i \sim \gig(a-1, 1, 2 |\theta_i|)$
    \\
    \qquad \enspace\qquad\qquad For $i=1$ to $i=n$
    \\
    \qquad \qquad\qquad\qquad Set $\phi_i = T_i / \sum_{h=1}^n T_h$
  \end{tabbing}
\end{algorithm}
While Step $1$ is a sample from the conditional distribution of $\theta$ given
$\psi, \tau, \phi,$ and $y$, Steps $2$--$4$ jointly are not a draw from the
distribution of $(\psi, \tau, \phi)$ given $\theta$. In fact, at Step $2$ the
components of $\psi$ are sampled conditionally on the current values of $\tau$
and $\phi$ and a similar problem arises for Step $3$. Critically, after
completing Step $4$, the new values of $\psi, \tau, \phi$ depend not only on the
current value of $\theta$ but also on the previous values of $\tau$ and $\phi$.
In fact, to obtain an independent sample from the desired full conditional
distribution, one samples $\phi|\theta$, then $\tau \mid \theta, \phi$, and
finally $\psi \mid \theta, \tau, \phi $. The wrong order is not a minor detail,
but it is something that leads to an incorrect Markov Chain Monte Carlo
procedure as discussed in the following proposition.
\begin{proposition}
  \label{prop:alg1fails}
  Let $q(\theta, \psi, \tau, \phi)$ be the stationary distribution of
  Algorithm \ref{alg:original}. Then
  \begin{equation*}
    q(\theta, \psi, \tau, \phi) \ne p(\theta, \psi, \tau, \phi \mid y).
  \end{equation*}
\end{proposition}
\begin{proof}
  See Appendix.
\end{proof}
Notably, the proof of Proposition 1 does not depend on the functional form of
$p(\theta \mid \psi, \tau, \phi, y)$ implying that the implications of
Proposition~\ref{prop:alg1fails} are general and not limited to the normal means
problem in \eqref{eq:nmp}, as also detailed in the following subsection.

We now provide a proper Gibbs sampler to perform Bayesian inference for model
\eqref{eq:nmp}. Trivially, it would be sufficient to permute the orders in which
the single updates of Algorithm \ref{alg:original} are presented. Additionally, while theoretically valid, the permuted order admits refinements
to enhance computational efficiency. Specifically, as also discussed in
\citet{bhattacharya2015}, reparameterizing $\delta_i = \tau \phi_i$ reveals that
$\delta_1, \ldots, \delta_n$ follow independent gamma distributions with shared
shape parameter $a$ and rate $1/2$. Consequently, the Dirichlet-Laplace prior
can be rewritten as
\begin{equation}
  \label{eq:dl_eg}
  \theta_i \mid \psi_i, \delta_i \sim \N(0, \psi_i \delta_i^2) \, , \,
  \psi_i \sim \Exp(1 / 2) \, , \, \delta_i \sim \Ga(a, 1 / 2) \, , \, i = 1,
  \dots, n \, .
\end{equation}
As a consequence, drawing from $p(\phi \mid \theta)$ and $p(\tau \mid \theta,
\phi)$ can be substituted with draws from $p(\delta \mid \theta)$, the
latter being a random vector with independent components whose densities are
\begin{equation*}
  p(\delta_i \mid \theta_i) \propto p(\theta_i \mid \delta_i) p(\delta_i)
  \propto \delta_i^{a - 2} \exp \Big\{ -\frac{1}{2} \Big( \delta_i + \frac{2
  |\theta_i|}{\delta_i} \Big) \Big\}.
\end{equation*}
The above expression is the kernel of the density of the $\gig(a-1, 1,
2|\theta_i|)$ which allows to define the following, more computational
efficient, Gibbs sampling algorithm.
\begin{algorithm}
  \label{alg:correct}
  Redundancy-free Gibbs sampler for Dirichlet-Laplace prior
  \begin{tabbing}
    \qquad \enspace \textit{Step} 1. Sample $\theta \mid \psi, \delta, y$
    \\
    \qquad \enspace\qquad\qquad For $i=1$ to $i=n$
    \\
    \qquad \qquad\qquad\qquad Set $\zeta^2_i = \{1 + 1 / (\psi_i
    \delta_i^2)\}^{-1}$
    \\
    \qquad \qquad\qquad\qquad Sample $\theta_i \mid \psi_i, \delta_i, y_i
    \sim \N(\zeta^2_i y_i, \zeta^2_i)$
    \\
    \qquad \enspace \textit{Step} 2. Sample $\delta \mid \theta$
    \\
    \qquad \enspace\qquad\qquad For $i=1$ to $i=n$
    \\
    \qquad \qquad\qquad\qquad Sample $\delta_i \mid \theta_i \sim \gig(a-1, 1, 2
    |\theta_i|)$
    \\
    \qquad \enspace \textit{Step} 3. Sample $\psi \mid \theta, \delta$
    \\
    \qquad \enspace\qquad\qquad For $i=1$ to $i=n$
    \\
    \qquad \qquad\qquad\qquad Sample $\widetilde{\psi}_i \mid \theta_i, \delta_i
    \sim \iG(\delta_i / |\theta_i|, 1)$ and set $\psi_i = 1 / \widetilde{\psi}_i$
  \end{tabbing}
\end{algorithm}

Notably, expression~\eqref{eq:dl_eg} and its implementation in
Algorithm~\ref{alg:correct} reveal two key insights about the Dirichlet-Laplace
prior. First, the Dirichlet distribution itself can be circumvented entirely in
posterior computation. Second, despite $\tau$ being shared across all
components, observing $\delta_i$ provides no information about $\delta_h$ ($h
\neq i$), as the product structure $\tau \phi_i$  preserves marginal
independence also \textit{a posteriori.}

\subsection{Linear Regression Model}
\label{subsec::lm}
We now consider a classical linear regression framework. Let $X$ denote the $n
\times p$ design matrix with entries $x_{ij}$, representing the $j$-th covariate
value for the $i$-th subject, where $j = 1, \dots, p$ and $i = 1, \dots, n$. The
response variables follow the linear model:
\begin{equation}
  \label{eq:lm}
  y_i = \sum_{j = 1}^p x_{ij} \theta_j + \varepsilon_i, \quad \varepsilon_i \overset{\text{i.i.d.}}{\sim} \N(0, \sigma^2).
\end{equation}
The Dirichlet-Laplace prior specified in \eqref{eq:DirLapl} admits the following
hierarchical representation for this regression framework:
\begin{equation}
  \label{eq:dl_lm}
  \begin{aligned}
    \theta \mid \sigma^2, \psi, \delta &\sim \N_p(0, \sigma^2 D),
  \end{aligned}
\end{equation}
where  $D$ is a diagonal matrix with $j$-th element in the diagonal equal to
$\psi_j \tau^2 \phi^2_j = \psi_j \delta^2_j$ where the right hand side of the
last equality employs the scale mixture representation presented in
\eqref{eq:dl_eg} for the hyperparameters $\psi_j$ and $\delta_j$. As customary
one can assume $\sigma^2 \sim \text{iGa}(s, r)$.

This approach was adopted by \citet{zhang2018}, who proposed an algorithm for
posterior computation, introducing an error analogous to the one discussed in
the previous section. In Algorithm \ref{alg:original_lm} the procedure of
\citet{zhang2018} is reported. Notably, the updates of $\theta$ and $\sigma^2$
have been collapsed into a single update of the joint full conditional
distribution of $(\theta, \sigma^2)$. Indeed, if $\sigma^2 \sim \text{iGa}(s,
r)$, the full conditional of $(\theta, \sigma^2)$ is normal-inverse gamma
distributed. This is a standard result that is omitted for brevity.
\begin{algorithm}
  \label{alg:original_lm}
  \citet{zhang2018} Gibbs sampler
  \begin{tabbing}
    \qquad \enspace \textit{Step} 1. Sample $(\theta, \sigma^2)  \mid  \psi,
    \tau, \phi, y$
    \\
    \qquad \enspace \textit{Step} 2. Sample $\psi \mid \theta, \sigma^2, \tau,
    \phi$
    \\
    \qquad \enspace\qquad\qquad For $j=1$ to $j=p$
    \\
    \qquad \qquad\qquad\qquad Sample $(\widetilde{\psi}_j \mid \theta_j,
    \sigma^2, \tau, \phi_j) \sim \iG(\sigma \tau \phi_j / |\theta_j|, 1)$ and set $\psi_j = 1 / \widetilde{\psi}_j$ 
    \\
    \qquad \enspace \textit{Step} 3. Sample $(\tau \mid \theta, \sigma^2, \phi)
    \sim \gig \big(p(a-1), 1, 2 \sum_{j=1}^p |\theta_j| / (\sigma \phi_j) \big)$
    \\
    \qquad \enspace \textit{Step} 4. Sample $\phi \mid \theta, \sigma^2$
    \\
    \qquad \enspace\qquad\qquad For $j=1$ to $j=p$
    \\
    \qquad \qquad\qquad\qquad Sample $T_j \sim \gig(a-1, 1, 2 |\theta_j| /
    \sigma)$
    \\
    \qquad \enspace\qquad\qquad For $j=1$ to $i=p$
    \\
    \qquad \qquad\qquad\qquad Set $\phi_j = T_j / \sum_{h=1}^p T_h$
  \end{tabbing}
\end{algorithm}
Similarly to Algorithm \ref{alg:correct}, the following scheme describes the
correct and redundancy-free algorithm to sample from the posterior distribution
of model \eqref{eq:lm}-\eqref{eq:dl_lm} exploiting $\delta_j = \tau \phi_j$ and
\eqref{eq:dl_eg}.
\begin{algorithm}
  \label{alg:correct_lm}
  Redundancy-free Gibbs sampler for linear model under Dirichlet-Laplace prior
  \begin{tabbing}
    \qquad \enspace \textit{Step} 1. Sample $(\theta, \sigma^2)  \mid  \psi, \delta, y$
    \\
    \qquad \enspace \textit{Step} 2. Sample $\delta \mid \theta, \sigma^2$
    \\
    \qquad \enspace\qquad\qquad For $j=1$ to $j=p$
    \\
    \qquad \qquad\qquad\qquad Sample $(\delta_j \mid \theta_j, \sigma^2) \sim
    \gig(a-1, 1, 2 |\theta_j| / \sigma)$ \\
    \qquad \enspace \textit{Step} 3. Sample $\psi \mid \theta, \sigma^2, \delta$
    \\
    \qquad \enspace\qquad\qquad For $j=1$ to $j=p$
    \\
    \qquad \qquad\qquad\qquad Sample $(\widetilde{\psi}_j \mid \theta_j,
    \sigma^2, \delta_j )\sim \iG(\sigma \delta_j / |\theta_j|, 1)$ and set $\psi_j = 1 / \widetilde{\psi}_j$
  \end{tabbing}
\end{algorithm}
In both schemes, samples from the full conditional distribution of $\theta$ can be obtained using the algorithm proposed by \citet{bhattacharya2016}, which is more efficient than the standard Cholesky decomposition of the conditional variance-covariance matrix when $p > n$.

\section{Numerical Experiments}
\label{sec:exp}

\subsection{Normal Means Problem}
\label{subsec::exp_nmp}
We now illustrate the practical consequences of using an incorrect Gibbs
sampler. Using a setting similar to \citet{bhattacharya2015}, we simulate data
under \eqref{eq:nmp} with $\theta^\dagger$ the true parameter values, $q_n$ the
number of non-zero entries of $\theta^\dagger$ and $A$ a non-zero constant. We
set $\theta^\dagger_i = A$ for $i\leq q_n$ and $\theta^\dagger_i = 0$ for $i >
q_n$. 

For each simulated data set, we (i) sample $y \sim \N_n(\theta^\dagger, I_n)$,
(ii) run Algorithms  \ref{alg:original} and \ref{alg:correct} for $20$,$000$
iterations, (iii) computed posterior medians as parameter estimates, and (iv)
calculate the squared error loss for these estimates. We vary $n \in \{100,
200\}$, $q_n / n \in \{ 0.05, 0.10, 0.20\}$, and $A \in\{ 5,6,7,8\}$ for a total
of $24$ different data generating processes. The global shrinkage hyperparameter
$a$ is fixed to $1/n$ or $1/2$. Each experiment is replicated $100$ times. 

\begin{table}[h!]
  \caption{Average squared loss for Algorithms \ref{alg:original} and \ref{alg:correct} with $n=100$.} 
  {
\setlength{\tabcolsep}{2.5pt}
\begin{tabular}{c c c | c c c c | c c  c c | c c c c}
    && $q_n /n$ & \multicolumn{4}{c}{0.05} & \multicolumn{4}{c}{0.10} & \multicolumn{4}{c}{0.20}\\
    & & $A$ & 5 & 6 & 7 & 8 & 5 & 6 & 7 & 8 & 5 & 6 & 7 & 8 \\
    \hline
    \multirow{4}{*}{$a$} & 
\multirow{2}{*}{$\frac{1}{n}$} & {\footnotesize Alg. \ref{alg:original}} &16.19 & 10.25 & 10.39 & 8.97 & 29.09 & 18.18 & 22.15 & 13.32 & 56.85 & 34.10 & 49.85&28.73\\
 & & {\footnotesize Alg. \ref{alg:correct}} &14.49 & 8.19 & 7.31 & 7.46 & 24.95 & 14.53 & 12.53 & 11.50 & 52.38 & 28.55 & 26.47&24.18\\
\cline{2-15}
& \multirow{2}{*}{$\frac{1}{2}$} & {\footnotesize Alg. \ref{alg:original}} &14.09 & 12.33 & 11.81 & 12.30 & 21.83 & 18.05 & 16.99 & 16.33 & 39.96 & 31.85 & 31.82&29.84\\
&  & {\footnotesize Alg. \ref{alg:correct}} &15.42 & 14.17 & 14.04 & 14.63 & 21.78 & 18.92 & 18.36 & 17.96 & 37.27 & 31.30 & 31.92&30.60\\
	\end{tabular}}
  \label{table:A5678}
\end{table}

Table \ref{table:A5678} reports the average of the squared error loss of the
replications for $n=100$. The results for $n=200$ are qualitatively similar and
reported in the Appendix. Notably, Algorithm \ref{alg:correct}
uniformly shows the best performance when \textit{a priori} the hyperparameter
$a$ is set to $1/n$. Algorithm \ref{alg:original} with $a = 1/2$ leads to a
lower squared error loss than Algorithm~\ref{alg:correct} under the same
hyperparameter choice. We suspect that $a = 1/2$ does not provide sufficient
shrinkage of those $\theta_i$'s whose $\theta^\dagger_i$'s are equal to zero.
Consistently, we recommend to use $a = 1/n$ as default choice. 

While evaluating the numerical performance of point estimates derived from the
algorithms in Section~\ref{sec:dirlapl} provides practical insights, it is
critical to emphasize that the output of Algorithm~\ref{alg:original} does not
constitute valid samples from the target posterior distribution. Consequently,
rigorous Bayesian uncertainty quantification must rely exclusively on outputs
generated by Algorithm~\ref{alg:correct}. Failure to do so risks propagating
erroneous conclusions, as any summary derived from Algorithm~\ref{alg:original}
lack theoretical guarantees.

\begin{center}
  \begin{figure}[h!]
    \caption{Logarithm of average squared loss for each $\theta_i$: Algorithm
    \ref{alg:original} (Original) on the $y$-axis vs Algorithm \ref{alg:correct}
    (Correct) on the $x$-axis. $\theta^\dagger_i = 8$ (top), $\theta^\dagger_i =
    0$ (bottom), $a = 1 / n$ (left), and $a = 1/ 2$ (right). $n = 200$,
    and $q_n = 40$. The red line is the identity function.}
    \centering
    \includegraphics[width=0.9\textwidth]{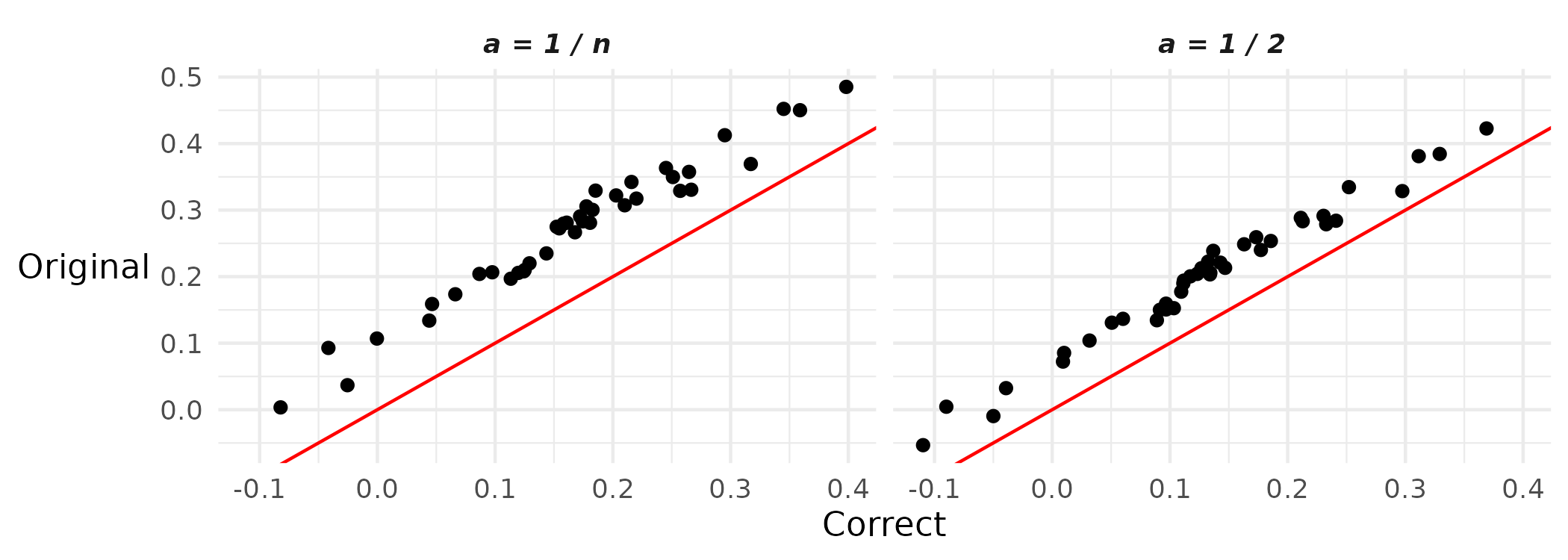}
    \includegraphics[width=0.9\textwidth]{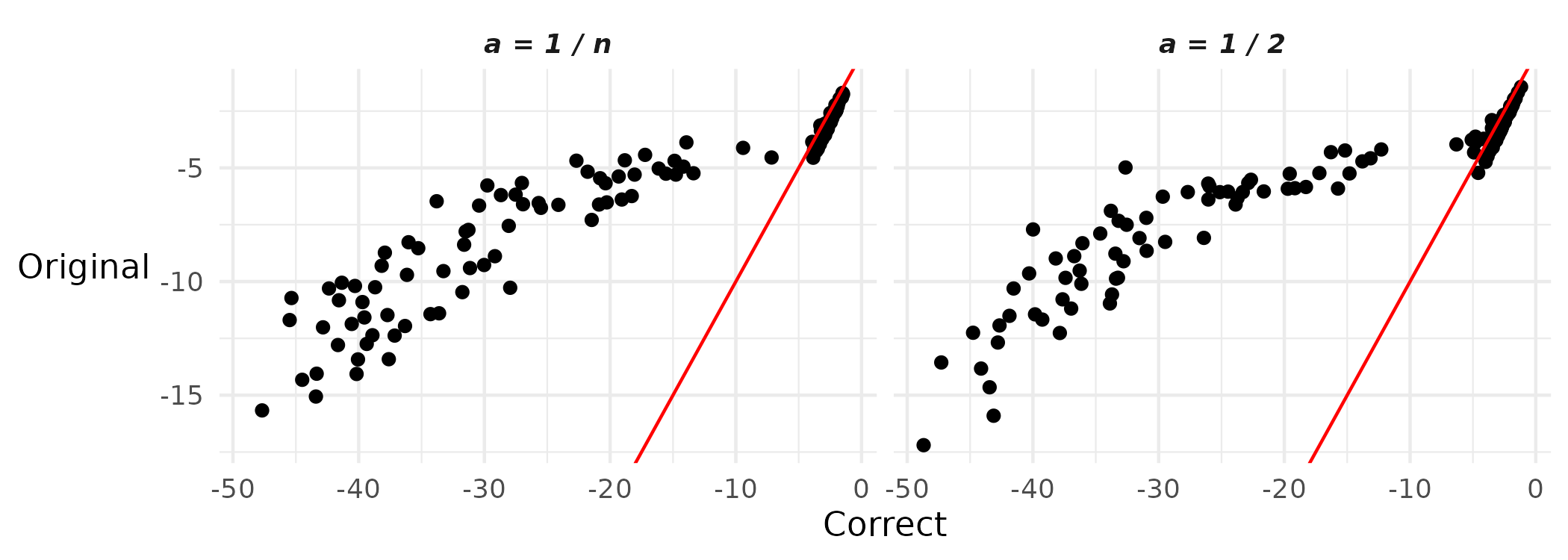}
    \label{fig:mse}
  \end{figure}
\end{center}

Figure \ref{fig:mse} compares the logarithm of the average squared loss for each
$\theta_i$ for the original flawed Gibbs sampler and the corrected one. This
shows that the correction produces better performance, with the gains
particularly pronounces for the non-zero signal coefficients. Additional results
are presented in the Appendix.

\subsection{Linear Regression Model}
\label{subsec::exp_lm}
We also provide results for linear regression, simulating 1,000 datasets from
\eqref{eq:lm} with $\sigma^2 =1$,  and true parameters:
\begin{align*}
  \theta^\dagger_1, \dots, \theta^\dagger_{p/2} = 0, \phantom{ \sigma}
  \qquad &
  \theta^\dagger_{p/2 + 2m + 1}, \dots, \theta^\dagger_{p/2 + 3m} = 7 \sigma,
  \\
  \theta^\dagger_{p/2 + 1}, \dots, \theta^\dagger_{p/2 + m} = 5 \sigma,
  \qquad &
  \theta^\dagger_{p/2 + 3m + 1}, \dots, \theta^\dagger_{p/2 + 4m} = 8 \sigma,
  \\
  \theta^\dagger_{p/2 + m + 1}, \dots, \theta^\dagger_{p/2 + 2m} = 6 \sigma,
  \qquad &
  \theta^\dagger_{p/2 + 4m + 1}, \dots, \theta^\dagger_{p/2 + 5m} = 10 \sigma,
\end{align*}
where $p = 100$,  $m = p / 10$, and  $n = 50$. Covariates are sampled
independently from a standard Gaussian distribution. The global shrinkage
hyperparameter $a$ is fixed to $1/p$, $1/n$, or $1/2$. For each replicate,
Algorithms \ref{alg:original_lm} and \ref{alg:correct_lm} are run for 20,000
iterations, and the squared error loss is computed.

Table \ref{table:lm} presents the results. The correct samplers consistently
improve performance in terms of average squared loss for both null and non-null
coefficients, regardless of the hyperparameter choice. However, the differences
diminish as the value of $a$ increases. Notably, smaller values of $a$ induce
stronger prior shrinkage, making the differences between the two algorithms more
pronounced when strong prior shrinkage is desired. This behavior can also be
observed in Figure~\ref{fig:mse_lm}, reported in the Appendix.
 \begin{table}[h!]
	\caption{Average squared loss for Algorithms \ref{alg:original_lm} and
   \ref{alg:correct_lm} for the null parameters (first column), non-null parameters (second column), and for both (third column).}
\begin{center} \begin{tabular}{ c  c | c | c c c}
    && & $\theta^\dagger_j = 0$ & $\theta^\dagger_j \neq 0$ & Total \\ 
    \hline
    \multirow{6}{*}{$a$} & 
		\multirow{2}{*}{$\frac{1}{p}$} & {\footnotesize Algorithm \ref{alg:original_lm}} & 817 & 2455 & 3273 \\
		               &                & {\footnotesize Algorithm \ref{alg:correct_lm}}  & 808 & 2324 & 3131 \\
\cline{2-6}
		&\multirow{2}{*}{$\frac{1}{n}$} & {\footnotesize Algorithm \ref{alg:original_lm}} & 831 & 2345 & 3176 \\ 
             &                      & {\footnotesize Algorithm \ref{alg:correct_lm}}  & 800 & 2315 & 3116 \\
\cline{2-6}
   & \multirow{2}{*}{$\frac{1}{2}$} & {\footnotesize Algorithm \ref{alg:original_lm}} & 478 & 2262 & 2740 \\ 
	&	                               & {\footnotesize Algorithm \ref{alg:correct_lm}}  & 471 & 2257 & 2729 
	\end{tabular} \end{center}
  \label{table:lm}
\end{table}

\appendix
\section*{Appendix}
\subsection*{Proof of Proposition 1}
\begin{proof}
\label{sec:proof}
Let $\thetac, \psic, \tauc, \phic$ be the current state of the
Markov chain and  $\thetan, \psin, \taun, \phin$ be the new values after a
complete scan.  The proof is based on showing that the posterior
distribution is not invariant with respect to the transition kernel of
Algorithm~\ref{alg:original}, defined as
\begin{equation*}
  \label{eq:kernel}
    \mathcal{K}(\thetac, \psic, \tauc, \phic \,;\,  \thetan, \psin, \taun,
    \phin) = p(\thetan \mid \psic, \tauc, \phic, y) p(\psin \mid \thetan, \tauc,
    \phic) p(\taun \mid \thetan, \phic) p(\phin \mid \thetan).
\end{equation*}

If $q(\theta, \psi, \tau, \phi)$ is equal to $p(\theta, \psi, \tau, \phi \mid
y)$, then the posterior distribution should be invariant with respect to the
transition kernel, or, in other words,  the following identity should hold true:
\begin{align}
  \label{eq:invarianceness}
  p(\thetan, \psin,& \taun, \phin \mid y)
  \\
  &= \int_{\thetac, \psic, \tauc, \phic}
  \mathcal{K}(\thetac, \psic, \tauc, \phic ;\, \thetan, \psin, \taun, \phin)
  p(\thetac, \psic, \tauc, \phic \mid y) d\thetac d\psic d\tauc d\phic \, . 
  \nonumber
\end{align}
We show that this is not the case. First we factorize the posterior on the left
hand side of \eqref{eq:invarianceness} as
\begin{equation}
  p(\thetan, \psin, \taun, \phin \mid y) = p(\thetan \mid y) p(\phin \mid
  \thetan) p(\psin, \taun \mid \thetan, \phin).
  \label{eq:factorizedposterior}
\end{equation}
Then we integrate the right hand side of equation above with respect to the
posterior distribution:
\begin{align}
  &\int_{\thetac, \psic, \tauc, \phic}
  \mathcal{K}(\thetac, \psic, \tauc, \phic ;\, \thetan, \psin, \taun, \phin)
  p(\thetac, \psic, \tauc, \phic \mid y) d\thetac d\psic d\tauc d\phic\notag
  \\
  =& \int_{\thetac, \psic, \tauc, \phic} p(\thetan \mid \psic, \tauc, \phic, y)
  p(\psin \mid \thetan, \tauc, \phic) p(\taun \mid \thetan, \phic) p(\phin \mid
  \thetan)\notag
  \\
  & \phantom{\int_{\thetac, \psic, \tauc, \phic}} p(\thetac, \psic, \tauc, \phic
  \mid y) d\thetac d\psic d\tauc d\phic\notag
  \\
  =& \int_{\psic, \tauc, \phic} p(\psin \mid \thetan, \tauc, \phic) p(\taun \mid
  \thetan, \phic) p(\phin \mid \thetan) p(\thetan, \psic, \tauc, \phic \mid y)
  d\psic d\tauc d\phic\notag
  \\
  =& \int_{\tauc, \phic} p(\taun \mid \thetan, \phic) p(\phin \mid \thetan)
  p(\thetan, \psin, \tauc, \phic \mid y) d\tauc d\phic\notag
  \\
  =& \int_{\phic} p(\taun \mid \thetan, \phic) p(\phin \mid \thetan)
  p(\thetan, \psin, \phic \mid y) d\phic \notag
  \\
  =& \, p(\phin \mid \thetan) \int_{\phic} p(\taun \mid \thetan, \phic)
  p(\thetan, \psin, \phic \mid y) d\phic.
  \label{eq:K}
\end{align}
Combining  \eqref{eq:invarianceness},  \eqref{eq:factorizedposterior}, and
\eqref{eq:K}, one gets
\begin{equation*}
  p(\thetan \mid y) p(\psin, \taun \mid \thetan, \phin) = \int_{\phic} p(\taun
  \mid \thetan, \phic) p(\thetan, \psin, \phic \mid y) d\phic, 
\end{equation*}
which is a contradiction because the right-hand side of the above formula does
not depend on $\phin$ and the identity must hold for every arbitrary value of
$\thetan, \psin, \taun, \phin$. Thus, the left-hand side must also not depend
on the specific value of $\phin$, that is, it must be a constant function with
respect to $\phin$. However, $p(\psin, \taun \mid \thetan, \phin)$ is available 
in closed-form and it is straightforward that is not constant with respect to 
$\phin$.
\end{proof}

\subsection*{Numerical Experiment: Additional Results}

Table~\ref{table:n200} reports  the average of the squared error loss mentioned
in Section~\ref{subsec::exp_nmp} for $n=200$.
Figures \ref{fig:nonzero} and \ref{fig:zero} show the distribution of the draws
of $\theta_{1}, \theta_{2}$ and of $\theta_{6}, \theta_{7}$, respectively, for a
single replicate of the experiment of Section \ref{subsec::exp_nmp} with $A =
8$, $n = 100$, and $q_n = 5$. In each panel we report a representation of the
20{,}000  draws where high density regions are colored proportionally to the
number of neighboring points using default options of the R package
\rpkg{ggplot2}{rpkg_ggplot2}. The crosses represent the true values of
$\theta^\dagger$. 
Specifically, Figure \ref{fig:nonzero} when $\theta^\dagger \ne 0$, shows that
Algorithm \ref{alg:original} over shrunk empirical distribution consistent with
results in \citet{bhattacharya2015} who concluded that $a=1/n$ produces over
shrinkage. Based on our findings, it appears this apparent over shrinkage was an
artifact of an incorrect sampler.
Conversely, Figure \ref{fig:zero} demonstrates that when the true signal is
zero, Algorithm \ref{alg:original} yields an empirical distribution that is less
concentrated around zero. This bias is particularly pronounced for $a = 1/n$
compared to $a = 1/2$.

\begin{center}
  \begin{figure}
    \caption{Logarithm of average squared loss for each $\theta_i$: Algorithm
    \ref{alg:original_lm} (Original) on the $y$-axis vs Algorithm
    \ref{alg:correct_lm} (Correct) on the $x$-axis. The red line is the identity
    function.}
    \includegraphics[width=\textwidth]{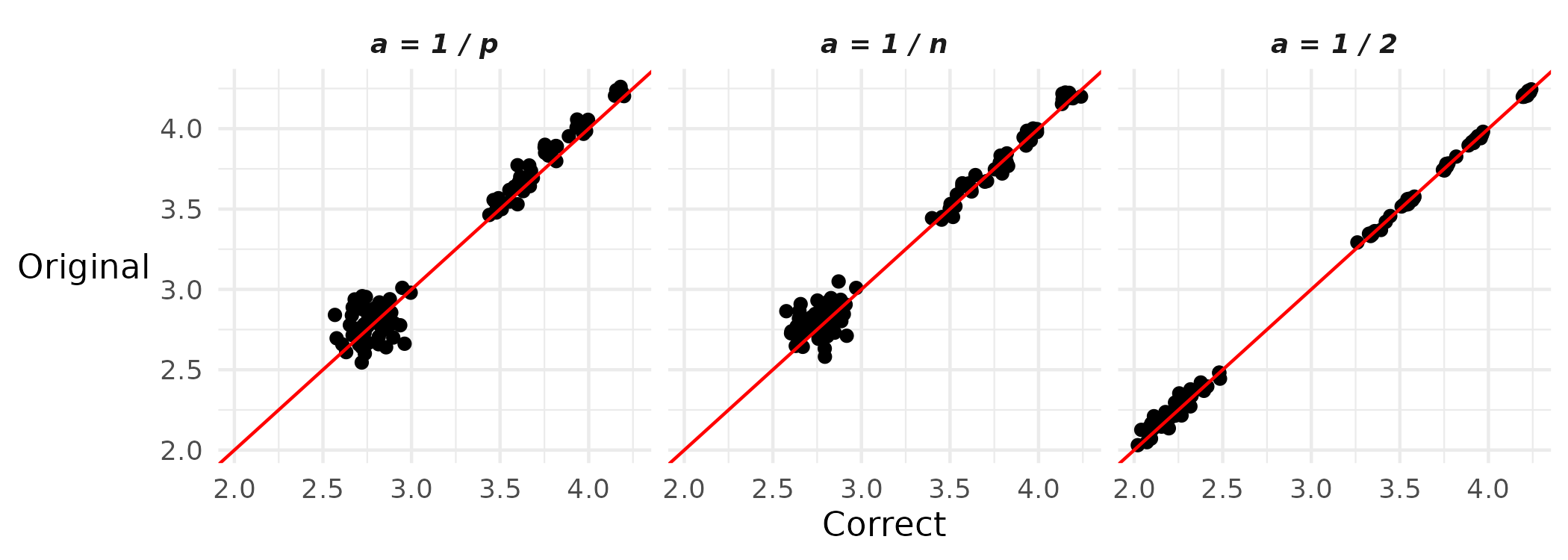}
    \label{fig:mse_lm}
  \end{figure}
\end{center}

Figure \ref{fig:mse_lm} displays the logarithm of the average squared error loss
for each $\theta_i$ when using Algorithm \ref{alg:original_lm}, plotted against
the corresponding value under Algorithm \ref{alg:correct_lm}.
\begin{table}[h]
  \caption{Average squared loss for Algorithms \ref{alg:original} and \ref{alg:correct} with $n=200$.} 
  {
\setlength{\tabcolsep}{4pt}
\begin{tabular}{c c c | c c c c | c c  c c | c c c c}
    && $q_n /n$ & \multicolumn{4}{c}{0.05} & \multicolumn{4}{c}{0.10} & \multicolumn{4}{c}{0.20}\\
    & & $A$ & 5 & 6 & 7 & 8 & 5 & 6 & 7 & 8 & 5 & 6 & 7 & 8 \\
    \hline
    \multirow{4}{*}{$a$} & 
\multirow{2}{*}{$\frac{1}{n}$} & {\footnotesize Alg. \ref{alg:original}} &30.63 & 20.48 & 24.78 & 18.38 & 52.89 & 38.30 & 49.60 & 30.81 & 107.65 & 70.71 & 97.46&61.41\\
&& {\footnotesize Alg. \ref{alg:correct}} &27.94 & 16.92 & 13.06 & 13.83 & 48.08 & 31.85 & 26.06 & 24.65 & 100.72 & 59.87 & 53.89&48.56\\
\cline{2-15}
& \multirow{2}{*}{$\frac{1}{2}$} & {\footnotesize Alg. \ref{alg:original}} &27.77 & 24.69 & 22.68 & 23.56 & 42.68 & 38.40 & 36.14 & 34.68 & 78.27 & 64.74 & 61.40&58.24\\
&  & {\footnotesize Alg. \ref{alg:correct}} &30.55 & 28.54 & 27.17 & 28.18 & 42.99 & 40.31 & 39.36 & 38.01 & 72.33 & 62.72 & 61.75&59.00\\
	\end{tabular}}
  \label{table:n200}
\end{table}
\begin{center}
  \begin{figure}
    \caption{Algorithm \ref{alg:original} (Original) vs Algorithm
    \ref{alg:correct} (Correct): $20 \, 000$ draws from $(\theta_1, \theta_2
    \mid y)$ with $n = 100$, $q_n = 5$ and $\theta^\dagger_1 = \theta^\dagger_2
    = 8$. Crosses represent true values of the parameters}
    \includegraphics[width=0.8\textwidth]{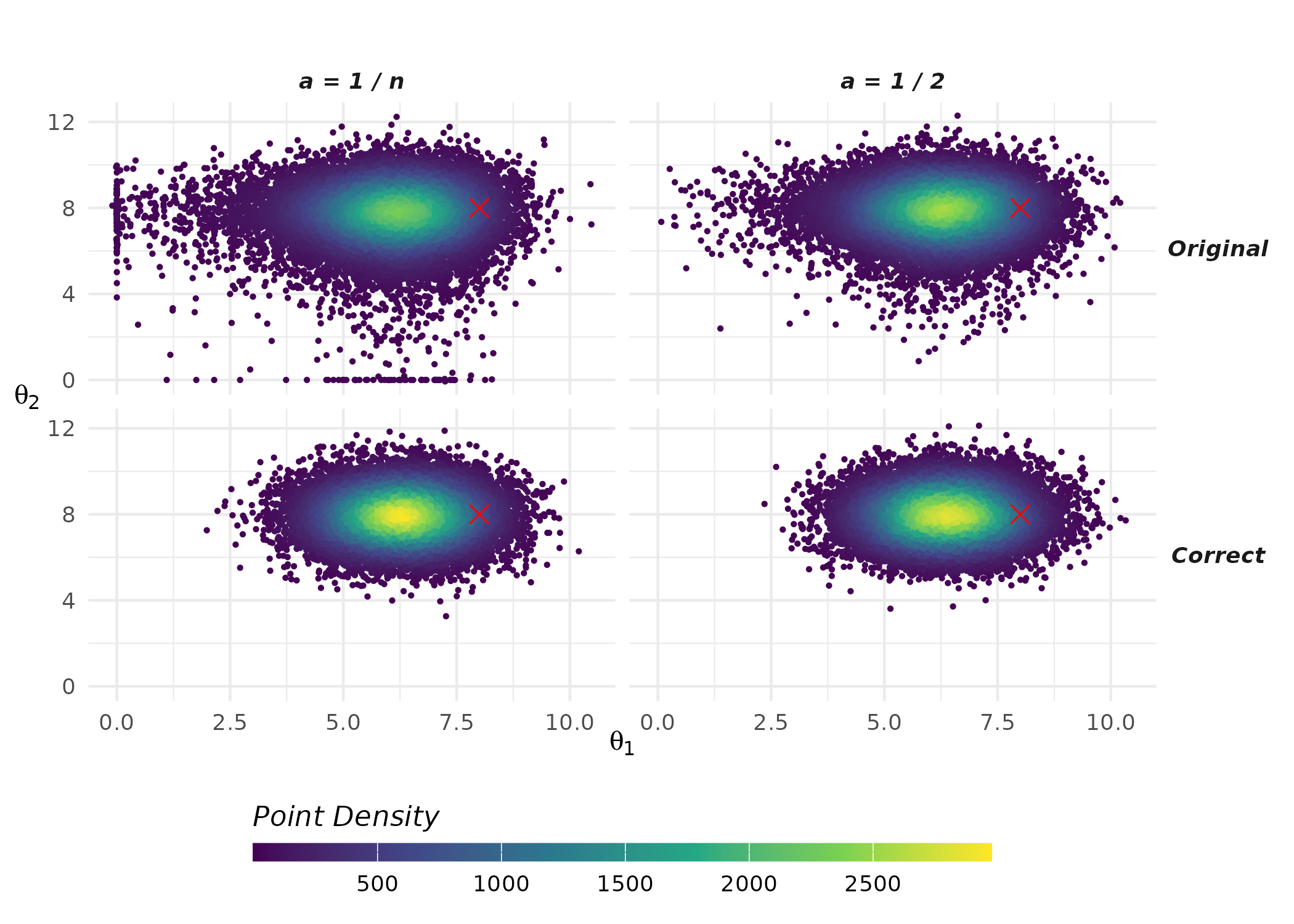}
    \label{fig:nonzero}
\end{figure}
\begin{figure}
    \caption{Algorithm \ref{alg:original} (Original) vs Algorithm
    \ref{alg:correct} (Correct): $20 \, 000$ draws from $(\theta_6, \theta_7
    \mid y)$ with $n = 100$, $q_n = 5$ and $\theta^\dagger_6 = \theta^\dagger_7
    = 0$. Crosses represent true values of the parameters}
    \includegraphics[width=0.8\textwidth]{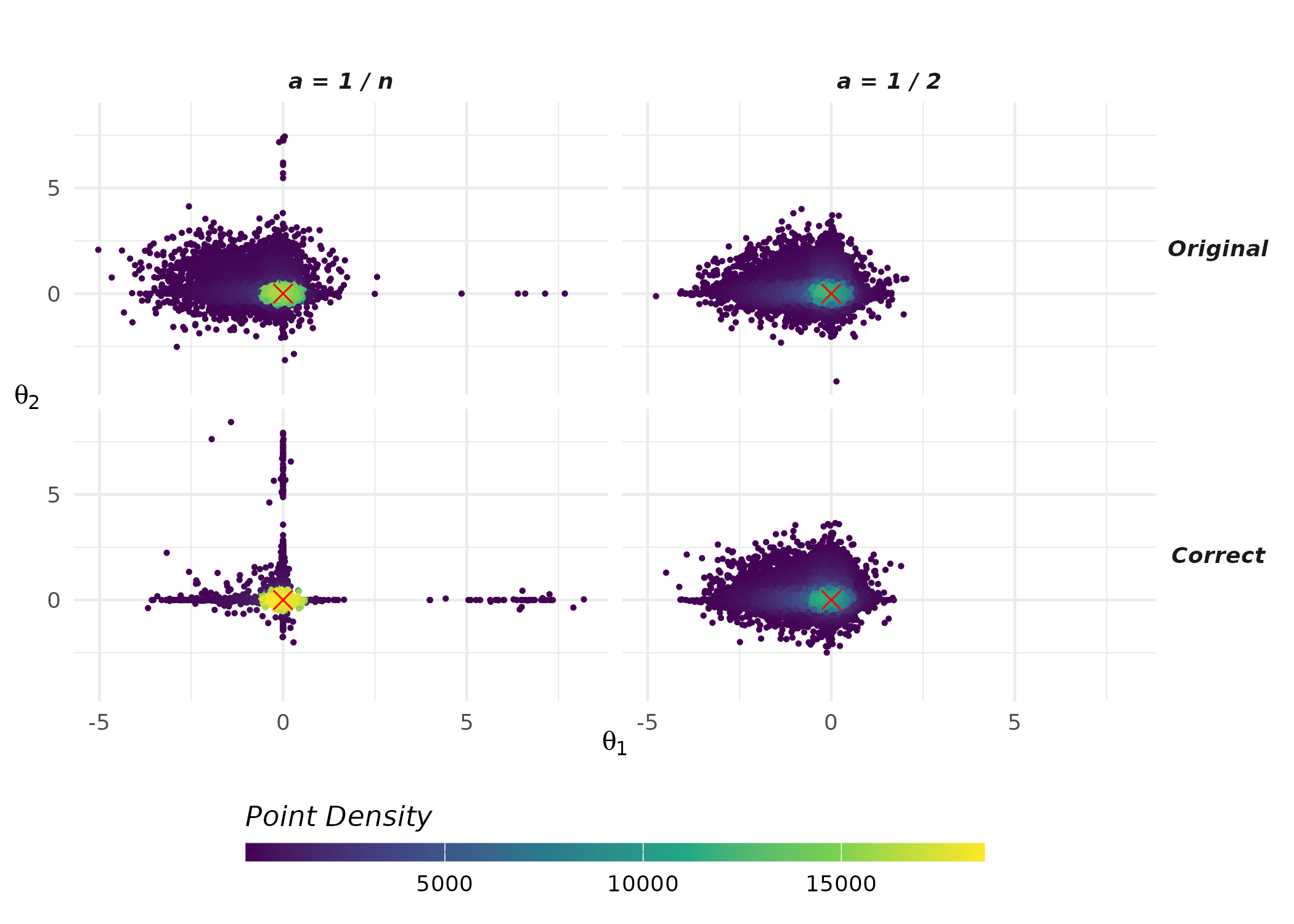}
    \label{fig:zero}
  \end{figure}
\end{center}

\section*{Supplementary Material}

\noindent
 An R package containing Gibbs sampling implementations for linear and logistic
 regression and an elliptical slice sampler for generic likelihoods is available
 at 
 \newline
 \textsc{github.com/paonrt/DirLapl}.

\section*{Acknowledgement}
\noindent
Paolo Onorati and Antonio Canale acknowledge support of MUR - Prin 2022 - Grant
no. 2022FJ3SLA, funded by the European Union – Next Generation EU. David Dunson
acknowledges support of the National Institutes of Health  grant R01ES03562.

\bibliography{idl.bib}

\end{document}